\documentclass[letterpaper, 10 pt, conference]{ieeeconf}  

\IEEEoverridecommandlockouts                              

\usepackage{amssymb}
\usepackage{amsmath}
\allowdisplaybreaks[1]

\usepackage{xcolor}
\usepackage{mathtools}
\usepackage{flushend}
\usepackage[]{algorithm2e}
\makeatletter
\renewcommand{\@algocf@capt@plain}{above}
\makeatother

\usepackage[protrusion=false]{microtype}

\newtheorem{theorem}{Theorem}
\newtheorem{definition}{Definition}
\newtheorem{lemma}{Lemma}
\newtheorem{proposition}{Proposition}
\newtheorem{corollary}{Corollary}

\let\sign\sgn

\DeclareMathOperator\Stab{Stab}
\DeclareMathOperator\Orb{Orb}

\def\P{\mathbb{P}}
\def\E{\mathbb{E}}
\title{Entropy Estimation of Physically Unclonable Functions
	via Chow Parameters}

\author{%
	\authorblockN{Alexander Schaub \qquad\qquad\qquad Olivier Rioul}
	\authorblockA{LTCI, Telecom Paris, Institut Polytechnique de Paris, France\\
		firstname.lastname@telecom-paristech.fr}
	\and
	\authorblockN{Joseph J. Boutros}
	\authorblockA{Texas A\&M University, 23874 Doha, Qatar\\
		boutros@tamu.edu}
}

\begin{document}

\maketitle

\begin{abstract}
	A physically unclonable function (PUF) is an electronic circuit that produces an intrinsic identifier in response to a challenge. These identifiers depend on uncontrollable variations of the manufacturing process, which make them hard to predict or to replicate. Various security protocols leverage on such intrinsic randomness for authentification, cryptographic key generation, anti-counterfeiting, etc.
	Evaluating the entropy of PUFs (for all possible challenges)  allows one to assess the security properties of such protocols. 
	
	In this paper, we estimate the probability distribution of certain kinds of PUFs composed of $n$ delay elements. This is used to evaluate relevant R\'enyi entropies and determine how they increase with $n$. Such a problem was known to have extremely high complexity (in the order of $2^{2^n}$) and previous entropy estimations were carried out up to $n=7$. 
	Making the link with the theory of Boolean threshold functions, we leverage on the representation by Chow parameters to estimate probability distributions up to $n=10$.
	The resulting Shannon entropy of the PUF is close to the max-entropy, which is asymptotically quadratic in $n$.
\end{abstract}


\section{Introduction}
Physically unclonable functions, or  PUFs, are electronic devices that
are  used to  produce  unique identifiers.   Small  variations of  the
manufacturing  process  are  exploited  so  that  any two  devices,  built
according to the same description, will likely produce different identifiers.
Moreover,  since such  process variations  are intrinsically  random,
they  cannot be  controlled  to  replicate  the behavior  of
another device,  hence the name \emph{physically unclonable} functions.
PUFs find many applications: the identifier can be
used  to generate  a  unique cryptographic  key,  which cannot  be
easily  extracted  from  the  device; it can be recorded during manufacturing
into a whitelist to prevent  counterfeiting or overproduction; and
it can also be employed in the  implementation of challenge-response
protocols at  a low  cost.
This is  especially valuable on devices where implementing asymmetric
cryptography primitives is too computationally expensive.

There are several ways to build PUFs. SRAM-PUFs~\cite{holcomb2009power}
exploit the states of SRAM  cells after powering up, while ring-oscillator (RO)
PUFs~\cite{suh2007physical} exploit delay differences of signals  in electronic circuits.  In this paper,  we analyze  another
delay  PUF,  called  loop-PUF,  first  proposed in~\cite{cherif2012easy}.  Our analysis  will also  be valid  for the
RO-sum PUF~\cite{yu2010recombination}, which shares  essentially the 
same mathematical model,   as    well   as   the   arbiter
PUF~\cite{gassend2003delay}. In the remainder of this paper, we will write
PUF as a short-hand for loop-PUF, RO-sum PUF or arbiter PUF.

\subsection{Modelization and Notations}
A  PUF  of   \textit{size}  $n$  generates  one   identifier  bit,  or
\textit{response}  bit,  when  queried with  a  \textit{challenge} 
$c =(c_1, \ldots, c_n)\in \left\{\pm 1\right\}^n$,  a
sequence of $n$ values $+1$ or $-1$. The PUF is characterized
by $n$ weights, denoted by $x  = (x_1, \ldots, x_n)  \in \mathbb{R}^n$
that represent  \textit{delay differences} of  the PUF circuit.  
As explained in~\cite{rioul2016entropy}, for each
challenge $c \in \left\{\pm 1\right\}^n$, the response bit of the PUF
of parameters $x  = (x_1, \ldots, x_n)$
is equal to 
$\sign(c\cdot x)=\sign(c_1x_1+\cdots+c_nx_n) \in \left\{\pm1\right\}$.

The base for all logarithms in this paper is equal to $2$, and all entropies are given in bits.

Due  to  manufacturing process  variations, the weights $x_i$ are 
modeled as realizations of random variables $X_i$. In~\cite{rioul2016entropy}, a Gaussian model was analyzed, where the Gaussian nature of the variables $X_i  \sim  \mathcal{N}(0,   1)$ is justified by simulations  of process  variations  in electronic
circuits~\cite{chang2003statistical}.

More generally, our analysis is valid for any $X=(X_1,X_2,\ldots,X_n)$ whose components $X_i$ are i.i.d. continuous variables with symmetric densities about $0$ (whose support contains $0$). 
The i.i.d. assumption is justified by the fact that delays are caused by ``identical'' circuit elements that lie in different locations in the circuit and can, therefore, be considered independent. In particular, each $x_i$ is the difference between two delays caused by such ``identical'' independent elements, which justifies the symmetry assumption.
Simulations in Section~\ref{sec:simulation} will be made in the Gaussian model, for which the weight distribution is centered isotropic.


\subsection{Problem Statement}

The  security  of  PUFs  is   related  to  R\'enyi  entropies
$H_\alpha$ of various orders~$\alpha$~\cite{renyi1961measures}.  

The min-entropy $H_\infty=\log (1/P_{\max})$ is related to the maximum (worst-case) probability $P_{\max}$ of successfully cloning a given PUF. Therefore, min-entropy $H_\infty$ should be as large as possible to ensure a given worst-case security level.

The collision entropy $H_2=\log(1/P_{\text{eq}})$ is related to the average probability $P_{\text{eq}}$ that two randomly chosen PUFs have the same identifier.
Therefore, $H_2$ should also be as large as possible to ensure a given average security level against collision.

The classical Shannon's entropy $H_1$ is known to provide a resistance criterion against modeling attacks---which predict the response to a new challenge given previous responses to other challenges~\cite{vijayakumar2016machine}. Again $H_1$ should be as large as possible.

The max-entropy $H_0$ is simply the logarithm of the total number of PUFs. $H_0$ upper bounds all the other entropies~$H_\alpha$. Theoretically, it is possible to choose a non i.i.d. weight distribution such that all PUFs are equiprobable, yielding $H_\alpha=H_0$ for every $\alpha$. In this case it is sufficient to count PUFs.
In practice, however, due to the assumption of i.i.d. weights (typically Gaussian),
the upper bound $H_0$ will not be attained. Therefore, it is important to derive a efficient method to estimate the various R\'enyi entropies. 

Estimating the various R\'enyi entropies typically requires estimating the entire 
PUF probability distribution. 
However, because  a PUF of size $n$ is  determined by $2^n$
response bits, there can be as many as $2^{2^n}$ PUFs of
size~$n$.  The  naive complexity increases very rapidly 
with $n$, in the order of $2^{2^n}$.

\subsection{Outline}
In this paper, we link the analysis of PUFs to the theory of Boolean Threshold Functions (BTF) and build an algorithm that accurately estimates the PUF  probability distribution and entropies up to order $n=10$.
Our algorithm relies on determining equivalent classes of PUFs with the same probability, and then estimating the probability within each class. The classes are determined using Chow parameters from BTF theory. The remainder of the paper is thus organized as follows. Section~\ref{sec:BTF} recalls known results from the theory of BTFs which we adapt to PUFs.  The key results on the equivalence classes are proved in Section~\ref{sec:eq_class}. Section~\ref{sec:algorithm} describes the simulation algorithm that allows us in Section~\ref{sec:simulation} to determine the PUF distributions and entropies up to order~$10$. Finally, Section~\ref{sec:conclusion} concludes.

\section{The Chow Parameters of PUFs}
\label{sec:BTF}
\begin{definition}[PUF]\label{def-puf}
	Let $x \in \mathbb{R}^n$ be such that for all $c \in \{\pm1\}^n$, $c\cdot x \neq 0$.
	The PUF of \emph{size} $n$ and \emph{weight} sequence $x$ is the function $f_x: \left\{-1, +1\right\}^n \to \left\{-1, +1\right\}$ defined as
	\begin{equation}
	f_x(c) = \sign(c\cdot x)
	\end{equation}
	where $c\cdot x= \sum_{i=1}^n{c_i x_i}$ is the usual scalar product.
\end{definition}

\medskip

This definition coincides with so-called ``self-dual'' BTFs of $n$ variables~\cite{goto1962some}. BTFs have been studied since the 1950's as building blocks for Boolean circuits~\cite{winder1961single} and also find applications in machine learning~\cite{cover1965geometrical}. Leveraging the correspondence between PUFs and BTFs, we adapt  fundamental results from BTF theory to conveniently characterize PUFs.

\subsection{All PUFs are Attainable}

Recall that in our framework, the PUF parameters $x\in \mathbb{R}^n$ are realizations of a random vector $X\in \mathbb{R}^n$. Under this probabilistic model a PUF becomes a randomized mapping $f_X$ such that $f_X(c)=\sign(c\cdot X)$ for any (deterministic) challenge $c \in \{\pm1\}^n$.

\begin{lemma}\label{lem:attainable}
	For every PUF $f_x$, we have $\P(f_X=f_x)> 0$.
\end{lemma}
In other words, every PUF $f_x$ can be reached by a realization of weights $X$ with positive probability (even though one has $\P(X=x)=0$).

\proof{
	By assumption all components of $X$ are i.i.d. with symmetric density of  support $S$ containing $0$.
	Hence the support $S^n$ of the density of $X$ is an $n$-dimensional manifold containing the origin in its interior.
	Let $x \in \mathbb{R}^n$ be fixed and let $C_x$ be the cone (scale-invariant set) of all $y\in\mathbb{R}^n$ such that $f_x=f_y$. This cone $C_x$ has apex $0$ and contains the intersection of all half-spaces $\{y\mid \sign(c\cdot y)= \sign(c\cdot x)\}$ where $c \in \{\pm1\}^n$. Therefore, it is a $n$-dimensional manifold which intersects  $S^n$ with positive volume. Hence $\P(f_X=f_x)=\P(X\in C_x\cap S^n)>0$.
}\endproof



\subsection{Chow Parameters Characterize PUFs}

First introduced by Chow~\cite{chow1961characterization} and later studied by Winder~\cite{winder1961single} who gave them their name, the so-called Chow parameters uniquely define a Boolean threshold function. Their definition is especially simple for PUFs:
\begin{definition}[Chow parameters]
	The \textit{Chow parameters} $p = (p_1, \dots, p_n)\in\mathbb{Z}^n$ of a PUF $f$ of size~$n$ is defined as
	\begin{equation}
	p = \!\!\!\sum_{c\mid f(c)=1}{c}	
	\end{equation}
	where the vector sum is carried out componentwise.
\end{definition}

\medskip

We remark that for $n\geq 2$, all Chow parameters are \emph{even integers}.
This is due to the fact that a sum of even number of elements $\pm1$ must be even.
More precisely, 
\begin{IEEEeqnarray}{ll}
	p_i \bmod 2\; & \equiv \sum_{c\mid f(c)=1}{c_i \bmod 2} \equiv \sum_{c\mid f(c)=1}{1} \bmod 2\\
	&\equiv 2^{n-1} \bmod 2 \equiv 0 \bmod 2.
\end{IEEEeqnarray}

\begin{theorem}[Chow's theorem~\cite{chow1961characterization}]
	\label{th_equal_pufs}
	Two PUFs with the same Chow parameters are identical.
\end{theorem}

\medskip

For completeness, we give a new proof of Chow's theorem rewritten in our PUF
framework. Such proof turns out to be very simple.

\proof{
	Let $f_x$ and $f_y$ be two PUFs with identical Chow parameters:
	\begin{equation}
	\smashoperator{\sum_{c\mid f_x(c)=1}}{\;c}\ = \smashoperator{\sum_{c\mid f_y(c)=1}}{\;c}.
	\end{equation}
	Simplifying this expression by $\displaystyle \smashoperator{\sum_{\substack{c\mid f_x(c)=1,\\ f_y(c)=1}}}{\;c}$, we obtain
	\begin{equation}
	\sum_{\substack{c\mid f_x(c)=1,\\ f_y(c)=-1}}{c}  ~~= \sum_{\substack{c\mid f_x(c)=-1,\\f_y(c)=1}}{c},
	\end{equation}
	which is equivalent to 
	\begin{equation}
	\smashoperator{\sum_{c\mid f_x(c)\neq f_y(c)}}{\;f_x(c)c}=0.
	\end{equation}
	Taking the scalar product with $x$, we get
	\begin{equation}
	\sum_{c\mid f_x(c)\neq f_y(c)}{f_x(c)c\cdot x} = \sum_{c\mid f_x(c)\neq f_y(c)}{{|c\cdot x|}} = 0
	\end{equation}
	which implies $c \!\cdot\! x = 0$ whenever $f_x(c)\neq f_y(c)$. Now we assumed that $c\cdot x$ is never zero by Def.~\ref{def-puf}. Thus $f_x=f_y$.
}\endproof

\subsection{Consequence on the Max-Entropy}	

An upper bound on the max-entropy can be easily deduced from Chow's theorem.

\begin{corollary}
	There are no more than $2^{n^2}$ PUFs of size $n$, i.e., the max-entropy of the PUF of size $n$ satisfies 
	\begin{equation}\label{eq:upperboundH0}
	H_0(n) \leq n^2 \qquad (\forall n \ge 2). 
	\end{equation}
\end{corollary}
A more refined version, which can be rewritten as $H_0(n)\leq (n-1)^2+1$ for $n>1$, can be found in~\cite[Corollary~10.2]{hu1965threshold}. The proof of~\eqref{eq:upperboundH0} is again particularly simple for PUFs. 

\medskip

\proof
	The Chow parameters $p_i$, $i=1 \ldots n$, satisfy 
	\begin{equation}
	p_i = \sum_{c\mid f(c)=1}{c_i} \leq \sum_{\substack{c\mid f(c)=1,\\c_i=1}}{1} \leq 2^{n-1}
	\end{equation}
	and similarly, $p_i \geq -2^{n-1}$. Since there are $2^{n-1}+1$ even integers between $-2^{n-1}$ and $2^{n-1}$,
	there can only be $(2^{n-1}+1)^n\leq 2^{n^2}$ different values taken by the Chow parameters.
	The conclusion follows from Chow's Theorem~\ref{th_equal_pufs}. 
\endproof

\medskip

A lower bound on $H_0$ is also easily found from the representation of Definition~\ref{def-puf}, as given by the following Proposition. The corresponding bound for the number of BTFs was first established independently by Smith~\cite{smith1966} and Yajima et al.~\cite{yajima1965} in the 1960s.

\begin{proposition}
	The max-entropy satisfies 
	\begin{equation}
	H_0(n) >  \frac{(n-2)^2}{2} \quad (\forall n\geq 2).
	\end{equation}
\end{proposition}

\medskip

\proof{
	Recall from Lemma~\ref{lem:attainable} that every PUF $f_x$ can be reached by a realization of weights $X$ with positive probability. Hence it is sufficient to consider all $f_x$ for all $x \in \mathbb{R}^n$ in order to lower-bound the total number of PUFs.

	Let $f_x$ a PUF of size $n$. Applying some small perturbation on $x$ if necessary (without affecting $f_x$) we may always assume that all the $c\cdot x$ ($c \in \left\{\pm 1 \right\}^n$) take distinct values. 
	
	Now let $x_{n+1} \in \mathbb{R}$ be such that $2x_{n+1}$ is different from all the $c\cdot x$, and define $x'=(x_1, \cdots, x_{n-1}, x_n-x_{n+1}, x_{n+1})$. For any challenge $c' = (c_1,\ldots, c_n, c_{n+1})$, we have 	
	\begin{equation}
	f_{x'}(c') = \begin{cases}
	f_x(c_1, \ldots, c_n)& \mbox{ if } c_n = c_{n+1}\\
	\sign(\sum_{i=1}^n{c_ix_i}-2c_n x_{n+1})& \mbox{ otherwise.}
	\end{cases}
	\end{equation}
	Depending on how many of the $2^{n-1}$ values of $c\cdot x$ are smaller/larger than $2c_nx_{n+1}$, we can construct $2^{n-1}+1$ different PUF functions of size $n+1$. Hence each PUF of size $n$ gives rise to more than $2^{n-1}$ PUFs of size $n+1$. Therefore, $H_0(n+1) > n-1 + H_0(n)$. The result follows by finite induction: $$H_0(n) > \frac{(n-1)(n-2)}{2} + H_0(2) > \frac{(n-2)^2}{2}.$$
}\endproof

\medskip

More recently, Zuev~\cite{zuev1992methods} has shown that, asymptotically, $H_0(n)>{n^2(1-\frac{10}{\ln(n)})}$. Therefore, for the max-entropy, we have that $H_0(n) \sim n^2$. 
As a result, instead of evaluating the probabilities of $2^{2^n}$ different PUFs, we will only have to evaluate about $2^{n^2}$.

As  apparent in the proof of Zuev~\cite[Theorem~1]{zuev1992methods} although through different geometrical considerations on normal vectors of hyperplanes,  we can further reduce the number of PUFs to be considered down by a factor of about ${2^nn!}$. Section~\ref{sec:eq_class} will derive the exact compression factor using the equivalence classes on Chow parameters.

\subsection{Order and Sign Stability of Chow Parameters}	

An important property of the Chow parameters $p$ is that  their share the same signs and relative order as the weights~$x$.
\begin{lemma}
	\label{lemma:chow-parameter-equivalence}
	Let $f = f_x$ be a PUF with weight $x\in\mathbb{R}^n$, and $p\in\mathbb{Z}^n$ be the corresponding Chow parameters. Then
	\begin{itemize}
		\item $x_i \geq 0 \implies p_i \geq 0$ and $x_i \leq 0 \implies p_i \leq 0$.
		\item $x_i \leq x_j \implies p_i \leq p_j$.
	\end{itemize}
\end{lemma}

\medskip

A similar result was shown by Chow in~\cite{chow1961characterization}, although with another definition of Chow parameters. Again we give a simplified proof in the PUF framework.

\proof{
	We first prove that $x_i \geq 0 \implies p_i \geq 0$, the other case $x_i \leq 0 \implies p_i \leq 0$ being similar. Suppose that $x_i \geq 0$.
	Let  $E_i^+$ (resp. $E_i^-$) be the set $\{c\mid f(c)=1, c_i =1\}$ (resp. $\{c\mid f(c)=1, c_i =-1\}$). By definition,
	\begin{equation}
	p_i = \sum_{c\mid f(c)=1}{c_i} = |E_i^+| - |E_i^-|.
	\end{equation}
	We show the existence of an injective mapping from $E_i^-$ to $E_i^+$.
	Consider the one-to-one mapping $\phi: \{\pm1\}^n \rightarrow \{\pm1\}^n$ defined by
	\begin{equation}
	\phi(c)_j =\begin{cases} 
	\hphantom{-}c_j, & j \neq i \\
	-c_j, & j=i
	\end{cases}
	\end{equation}
	For any $c\in E_i^-$,  $c_i=-1$, $\phi(c)_i=+1$ and
	\begin{IEEEeqnarray}{ll}
		\sum_{j=1}^n{\phi(c)_jx_j} &= \sum_{j\neq i}^n{c_jx_j} + x_i
		\\
		&= \underbrace{\sum_{j=1}^n{c_jx_j}}_{>0} + \underbrace{2x_i}_{\geq 0} > 0.
	\end{IEEEeqnarray}
	Therefore, $f(\phi(c)) =1$ and $\phi(c) \in E_i^+$. Hence, the bijection~$\phi$ induces an injection
	from $E_i^-$ to $E_i^+$. This implies that $|E_i^+| \geq |E_i^-|$ hence $p_i \geq 0$.
	
	To prove the second part, assume that $x_i \le x_j$ for $j \ne i$.
	Let $f': \{\pm1\}^{n-1} \rightarrow \{\pm1\}$ be a PUF given by $f'(c')=\sign(c'\cdot x')$,
	where $c'\in \{\pm1\}^{n-1}$ is obtained from $c$ by dropping $c_i$, $x'_{\ell}=x_{\ell}$
	for any $\ell \ne j$, and $x'_j=x_j-x_i \ge 0$. Say the Chow parameters of $f'$ is $p'$.
	According to the first part of this lemma, we have $p'_j \ge 0$. 
	Now, expand the expression of $p_j-p_i$ as
	\begin{IEEEeqnarray}{ll}
		p_j-p_i&=\sum_{c\mid f(c)=1}{c_j}-\sum_{c\mid f(c)=1}{c_i}\\
		&=2 \sum_{c\mid f(c)=1,c_j=-c_i}{c_j}\\
		&=2 \sum_{c'\mid f'(c')=1}{c'_j}=2p'_j\ge0. \quad
	\end{IEEEeqnarray}
}\endproof

\section{Equivalence Classes and Chow Parameters}
\label{sec:eq_class}
Since the $X_i$ are i.i.d. symmetric random variables,
the joint probability distribution of the weights $X=(X_1, \dots, X_n)$ is invariant under permutations and sign changes. Therefore, all PUFs $f_x$ that can be obtained from one another by permuting or changing signs of their weights $x_1,x_2,\ldots,x_n$ can be clustered together into equivalence classes of PUFs with the same probability $\mathbb{P}(f_X = f_x)$. 

We now establish several properties of these equivalence classes for PUFs, known as ``self-dual'' classes~\cite{goto1962some} in the context of BTFs.
Zuev~\cite{zuev1992methods} had already mentioned $2^n n!$ elements per class in a special case. Our generalization (Theorem~\ref{thm:orbite}) is mentionned in a different form in~\cite[\S~3.1.2]{gruzling2008linear} for calculating the total number of BTFs, yet we couldn't find formal proofs published in the literature.

We give a formal definition of the equivalence classes by the action of the group 	\begin{equation}
G_n = S_n \times \{-1, +1\}^n
\end{equation}
where $S_n$ is the symmetric group of order $n!$.
An element $g = (\sigma, s) \in G_n$ is determined by the permutation $\sigma\in S_n$ and the sign changes $s\in \{-1, +1\}^n$.

\smallskip

\begin{proposition}
	For any $x=(x_1, \dots, x_n) \in \mathbb{R}^n$ and $g=(\sigma, s)\in G_n$ define $g\cdot x : G_n \times \mathbb{R}^n \to \mathbb{R}^n$ such that
	\begin{equation}
	(g\cdot x)_i = s_ix_{\sigma(i)}.
	\end{equation}
	This defines a group action of $G_n$ on $\mathbb{R}^n$, where
	the inner product in $G_n$ is defined by
	\begin{equation}
	(\sigma_1, s^1) \cdot (\sigma_2, s^2) = (\sigma_1 \circ \sigma_2, (s^1_is^2_{\sigma_1(i)})_i).
	\end{equation}
\end{proposition}

\medskip

\proof{
	$G_n$ is clearly a group with identity $e = (id, (1,\cdots,1))$.
	For any $(\sigma_1, s^1), (\sigma_2, s^2) \in G_n$ and $x \in \mathbb{R}^n$,
	\begin{IEEEeqnarray}{ll}
		(\sigma_1, s^1) \cdot ((\sigma_2, s^2)\cdot x) &= (\sigma_1, s^1) \cdot (s^2_ix_{\sigma_2(i)})_i\\
		& = (s^1_is^2_{\sigma_1(i)}x_{\sigma_1(\sigma_2(i))})_i\\
		&=  (\sigma_1 \circ \sigma_2, (s^1_is^2_{\sigma_1(i)})_i) \cdot x\\
		& = ((\sigma_1, s^1) \cdot (\sigma_2, s^2))\cdot x.
	\end{IEEEeqnarray}
	This shows that $g\cdot x$ defines a group action of $G_n$ on $\mathbb{R}^n$.
}\endproof

\medskip

Thus we can say that the group $G_n$ acts on the PUFs of size~$n$, the action being defined as 
\begin{equation}
g\cdot f_x = f_{g\cdot x}. 
\end{equation}
In keeping with Lemma~\ref{lemma:chow-parameter-equivalence}, we now show that the group action is carried over to Chow parameters:


\begin{theorem}
	\label{thm:group-action-equivalence}
	Let $f_x$ a PUF of Chow parameters $p$, and let $g \in G_n$. The Chow parameters of $f_{g\cdot x}$ is $g \cdot p$.
	
\end{theorem}

\proof{
	Let $g = (\sigma, s)\in G_n$. For any challenge $c$, we have that $f_x(g^{-1}\cdot c) = f_{g\cdot x}(c)$. Thus,
	\begin{IEEEeqnarray}{ll}
		\sum_{c\mid f_{g\cdot x}(c) = 1}\!\!\!{c_i} ~& = \sum_{c\mid f_x(g^{-1}\cdot c)=1}\!\!\!\!\!{c_i} = \sum_{c\mid f_x(c)=1}\!\!\!{(g\cdot c)_i} \\
		&=\sum_{c\mid f_x(c)=1}\!\!\!\!{s_ic_{\sigma(i)}}	= s_ip_{\sigma(i)} = (g\cdot p)_i.  
	\end{IEEEeqnarray}
	
}
\endproof

\medskip

Changing the signs of the weights or permuting them is reflected by the same operation on the Chow parameters. This allows us to compute the size of the equivalence classes:

\smallskip

\begin{theorem}\label{thm:orbite}
	Let $f$ be a PUF with Chow parameters $p$. Let $m_p(k)$ be the number of Chow parameters equal to $k\in\mathbb{Z}$, and let $\Orb(f) = \{g\cdot f \mid g\in G_n \}$ the \textit{orbit} of $f$ by $G_n$, that is, the equivalence class containing $f$. Then
	\begin{equation}
	|\Orb(f)| = 2^n n! \Bigl(2^{m_p(0)}\prod_{k\in\mathbb{Z}}{m_p(k)!}\Bigr)^{-1}.
	\end{equation}
\end{theorem}

 \proof{
	By applying the well-known \textit{orbit-stabilizer theorem} (see for instance~\cite[p.~89]{hungerford1980algebra}),  we have 
	\begin{equation}
	\!\!\!\! |\Orb(f)| = \frac{|G_n|}{|\Stab(f)|} = \frac{ |\{\pm1\}^n| \times |S_n|}{|\Stab(f)|} = \frac{2^nn!}{|\Stab(f)|}
	\end{equation}
	where $\Stab(f) = \{g \in G_n \mid g\cdot f = f \}$ is the \textit{stabilizer} of $f$.
	The size of the orbit of $f$ can therefore be deduced from the size of its stabilizer. Now the latter can be easily computed: Let $g =(\sigma, s)\in G_n$ such that $g\cdot f = f$. Since $g\cdot p = p$, we have $\sigma(i) = j \iff p_i=p_j$ and $s_i = -1\iff p_i=0$. The number of such $g$ is exactly $2^{m_p(0)}\prod_{k\in\mathbb{Z}}{m_p(k)!}$. 
}\endproof

\section{Monte-Carlo Algorithm}\label{sec:algorithm}

As seen in the introduction to the previous section, all PUFs in one equivalence class have the same probability. It follows that the probability of any particular PUF can be deduced from the probability of the class to which it belongs. Therefore, to determine  the various entropies, it suffices to find a method that estimates the probabilities of the various equivalence classes. 

In this section, we propose an algorithm that exploits a definition of a \textit{canonical} PUF in any equivalence class in such a way that for given any PUF, it is trivial to determine the corresponding canonical PUF. As expected, only about  ${2^{n^2}}/{2^nn!}$ probabilities need to be estimated,  instead of approximatively~$2^{n^2}$. 

\smallskip

\begin{definition}[Canonical PUF]
	A \textit{canonical} PUF of $n$ variables is a PUF whose Chow parameters satisfy
	\begin{equation}
	p_1 \geq p_2 \geq \cdots \geq p_n \geq 0.
	\end{equation}
	The \textit{canonical form} of a PUF $f$ is the canonical PUF belonging to the same class, i.e., $f' = g\cdot f$ where $g\in G_n$ is such that $f'$ is canonical.
\end{definition}

This notion was first introduced by Winder~\cite{winder1961single} and is related to the concept of ``{prime}'' functions independently studied by Chow~\cite{chow1961characterization}.

\begin{proposition}[Unicity of the canonical PUF]
	Two canonical PUFs in the same class are equal.
\end{proposition}

\proof{
	Since $f$ and $f'$ are in the same equivalence class, their Chow parameters are identical up to sign changes and order. Since both are canonical, the signs and order are fixed. Their Chow parameters are thus identical and $f=f'$.
}\endproof

\medskip

\begin{proposition}
	Let $x = (x_1,\ldots,x_n)$ be a weight sequence of a PUF $f=f_x$, and let $g \in G_n$ such that $g\cdot x = (x'_1,\ldots x'_n)$ satisfies
	\begin{equation}
	x'_1 \geq x'_2 \geq \ldots \geq x'_n \geq 0.
	\end{equation}
	Then $g\cdot f$ is the canonical form of the PUF $f$.
\end{proposition}

\proof{
	Let us denote by $p$ (resp $p'$) the Chow parameters of $f$ (resp $g\cdot f$). The PUF obtained from weights $x'$ is $g\cdot f$. From Lemma~\ref{lemma:chow-parameter-equivalence}, the $p'_i$ satisfy the same ordinal relations and have the same signs as the $x'_i$. Therefore, $f'$ is a canonical~PUF.
}\endproof

\medskip

These results allow us to efficiently estimate the PUF distribution by Monte-Carlo methods, as described in Algorithm~\ref{alg:simulation}. Such an algorithm can  be used for any i.i.d. weight distribution with symmetric densities (not necessarily Gaussian).

\begin{algorithm} \small

\caption{How to estimate the PUF distribution.}
\label{alg:simulation}

	\SetKwData{Counts}{counts}\SetKwData{Proba}{proba}\SetKwData{Size}{size}
	\KwData{$n > 0, nbRounds > 0$}
	\KwResult{Estimation of PUF probability distribution }
	Initialize HashMaps \Counts, \Proba, \Size \;
	\For{$i\leftarrow 1$ \KwTo $nbRounds$}{
		Generate $n$  realizations $x_1, \ldots, x_n$\;
		Sort the absolute values of the $x_i$ to obtain $x'$\;
		Compute the Chow parameters $p$ of $f_{x'}$\;
		\eIf{$p \in$ \Counts}{
			\Counts[$p$] $\leftarrow$ \Counts[$p$] + 1\;
		}{
		\Counts[$p$] $\leftarrow 1$\;
	}
}
\For{$p\in$ \Counts}{
	$\displaystyle \Size[p] \leftarrow \frac{2^n n!}{2^{m_p(0)}\prod_{k}{m_p(k)!}}$ \;
	$\displaystyle \Proba[p] \leftarrow \frac{\Counts[p]}{\Size[p]*nbRounds}$\;
}
\Return (\Proba, \Size) \; \medskip

\end{algorithm}

\section{Entropies Estimation}\label{sec:simulation}

In this section, we present the simulation results in the Gaussian case where the weights $X_i$ are i.i.d. $\sim \mathcal{N}(0,1)$.
Exact values were already determined up to $n=4$ in~\cite{schaub2018challenge}.

\subsection{Estimating the Max-Entropy $H_0$}
According to Lemma~\ref{lem:attainable}, every PUF can be attained by some realization of weights. Therefore, the max-entropy of the PUF distribution is simply the logarithm of the total number of PUFs with $n$ weights. This number is equal to the total number of BTFs of $n-1$ variables and has been computed up to $n=10$ in~\cite[\S~3.1.2]{gruzling2008linear}, see Table~\ref{tbl:max-entropy}.

\begin{table}[htb!]\centering
	\caption{Exact values of $H_0$}
	\label{tbl:max-entropy}
	\begin{tabular}{|c|c|c|}
		\hline
		$n$  & \# PUFs & $H_0$ (bits) \\
		\hline
		1 & 2 & 1\\
		2 & 4 & 2\\
		3 & 14 & 3.8074\ldots\\
		4 & 104 & 6.7004\ldots\\
		5 & 1882 & 10.8781\ldots\\
		6 & 94572 & 16.5291\ldots\\
		7 & 15028134 & 23.8411\ldots\\
		8 & 8378070864 & 32.9640\ldots\\
		9 & 17561539552946 & 43.9974\ldots\\
		10& 144130531453121108 & 57.0001\ldots\\
		\hline
	\end{tabular}
\end{table}

\subsection{Estimating the Shannon Entropy $H_1$}

For any PUF $f$, let  $[f]$ denote the equivalence class of~$f$ with cardinality $|[f]|$, $\P(f)$ its probability, $F_n$ the set of all PUFs  and $F_n/G_n$ the quotient group induced by the action of the group $G_n$. Then, letting $\P([f']) = \sum_{f\in [f']}{\P(f)}$, one has
\begin{align}
H_1(n)\!&= -\!\sum_{f\in F_n} \P(f)\log(\P(f))
\\
&= -\!\!\sum_{f'\in F_n/G_n} \sum_{f\in [f']} \P(f)\log(\P(f))\\
&= -\!\!\sum_{f'\in F_n/G_n} \P([f'])\log(\P(f'))\\
&= -\!\!\sum_{f'\in F_n/G_n}{\P([f'])\log(\P([f']))} 
+ \mathbb{E}[\log(|[f_X]|)].
\end{align}
In other words, the Shannon entropy of the PUF distribution is simply the sum of the entropy of the equivalence classes and the average of their logarithmic size.
The latter term can be estimated using the unbiased empirical mean, where a confidence interval can be determined using Student's t-distribution~\cite{student1908probable}. The former term, however, is an entropy, for which no unbiased estimator exists~\cite{paninski2003estimation}. 
The NSB estimator~\cite{nemenman2002entropy} has a reduced bias and a low variance. However, because we generated much more PUFs than equivalence classes (by a factor of at least 100000), the plug-in estimator, based on the empirical frequency estimates, performs quite well: Its bias can be upper	 bounded as described in~\cite{paninski2003estimation} and was found to be less than $0.01$ bit.
The results are summarized in Table~\ref{tbl:shannon-entropy}.

\begin{table}[h!]\centering
	\caption{{Confidence intervals at the 95\% level for $H_1$\hspace{\linewidth} (exact values up to $n=4$).}}
	\label{tbl:shannon-entropy}
	\begin{tabular}{|c|c|c|}
		\hline
		$n$ & $\mbox{PUF Sample size}$  & $H_1$ (bits) \\
		\hline
		1 & --- & 1 \\
		2 & --- & 2 \\
		3 & --- & 3.6655\ldots \\
		4 & --- & 6.2516\ldots \\
		5 & $10^{10}$ & 10.0134 -- 10.0156\\
		6 & $10^{10}$ & 15.1903 -- 15.1925\\
		7 & $10^{10}$ & 21.9856 -- 21.9879\\
		8 & $2\cdot 10^{10}$ & 30.5628 -- 30.5645\\
		9 & $2\cdot 10^{10}$ & 41.0367 -- 41.0384\\
		10& $3\cdot 10^{12}$  & 53.4737 -- 53.4740\\
		\hline
	\end{tabular}
\end{table}


\subsection{Estimating the Collision Entropy $H_2$}

The collision entropy was estimated using an unbiased estimator adapted from~\cite[\S~1.4.2]{acharya2014complexity}. Let $N_{[f]}$ be the number of PUF samples that belong to the equivalence class of $[f]$ among a number of Poisson-distributed PUFs with parameter $N$, and  $N_{[f]}^{\underline{2}} = N_{[f]}\cdot(N_{[f]}-1)$. 
We can compute 
\begin{align}
	\E\Bigl[\sum_{f\in F_n/G_n} \frac{N_{[f]}^{\underline{2}}}{|[f]|N^2}\Bigr] &= \sum_{f\in F_n/G_n} \E\Bigl[\frac{N_{[f]}^{\underline{2}}}{N^2}\Bigr] \frac{1}{|[f]|}\\
	&= \sum_{f\in F_n/G_n} \frac{\P([f])^2}{|[f]|} \\
	&= \sum_{f\in F_n}{\P(f)^2}
\end{align}
where we used the fact that $\E[\frac{N_{[f]}^{\underline{2}}}{N^2}] = \P([f])^2$ from~\cite[\S~2.2]{acharya2014complexity}.
It follows that 
\begin{align}
\sum_{f\in F_n/G_n} \frac{N_{[f]}^{\underline{2}}}{|[f]|N^2}.
\end{align}

is an unbiased estimator for the power-sum $\sum_{f\in F_n}{\P(f)^2}$.
As can be also checked, the variance of this estimator admits the same upper bound 
as the one described in~\cite[\S~1.4.2]{acharya2014complexity}. This allows us to determine confidence intervals for the collision entropy as shown in Table~\ref{tbl:collision-entropy}.

\begin{table}[h!]\centering
	\caption{Confidence intervals at the 95\% level for $H_2$\hspace{\linewidth}(exact values up to $n=4$)}
	\label{tbl:collision-entropy}
	\begin{tabular}{|c|c|c|}
		\hline
		$n$ & $\mbox{PUF Sample size}$ & $H_2$ (bits) \\
		\hline
		1 & ---  & 1\\
		2 & --- & 2\\
		3 & --- & 3.5462\ldots\\
		4 & --- & 5.7105\ldots\\
		5 & $10^{10}$ & 8.4551 -- 8.4568 \\
		6 & $10^{10}$ & 11.5977 -- 11.6023 \\
		7 & $10^{10}$ & 14.8819 -- 14.89805 \\
		8 & $2\cdot 10^{10}$ &18.5201 -- 18.5753\\
		9 & $2\cdot 10^{10}$ & 22.0309 -- 22.4067\\
		10& $3\cdot 10^{12}$ & 25.9070 -- 26.1983\\
		\hline
	\end{tabular}
\end{table}

\subsection{Estimating the Min-Entropy $H_\infty$}

In order to determine the min-entropy of the PUF distribution, one needs to estimate the probability of the most likely PUF. Our experiments, as well as those of Delvaux et al.~\cite{delvaux2016upper}, strongly suggest that for a Gaussian distribution of the weights, the most likely PUFs are the $2n$ PUFs corresponding to the Boolean functions $c_i$ and $\overline{c_i}$, $i = 1...n$.

The maximum likelihood estimator of that probability is simply the sample frequency, which is an unbiased estimator. A confidence interval for this estimator can be obtained using the Wilson score interval~\cite{wilson1927probable}, which yields a confidence interval for the min-entropy $H_\infty$.

Because we have already determined that there are exactly $2n$ PUFs in the equivalence class of the most likely PUF, we only need to estimate a confidence interval on the sample frequency of the \textit{equivalence class}. Once such an interval was obtained, for instance $[p_-, p_+]$, then the confidence interval for the min-entropy is given by
$$[-\log_2(p_+)+\log_2(2n), -\log_2(p_-)+\log_2(2n)].$$
The confidence intervals of the min-entropy are presented in Table~\ref{tbl:min-entropy}.

\begin{table}[h!]\centering
	\caption{Confidence intervals at the 95\% level for $H_\infty$\hspace{\linewidth}(exact values up to $n=4$)}
	\label{tbl:min-entropy}
	\begin{tabular}{|c|c|c|}
		\hline
		$n$ & $\mbox{PUF Sample size}$ & $H_\infty$ (bits) \\
		\hline
		1 & ---  & 1\\
		2 & --- & 2\\
		3 & --- & 3.2086\ldots\\
		4 & --- & 4.5850\ldots\\
		5 & $10^{10}$ & 6.1006 -- 6.1008 \\
		6 & $10^{10}$ & 7.7352 -- 7.7354\\
		7 & $10^{10}$ & 9.4731 --  9.4735 \\
		8 & $2\cdot 10^{10}$ & 11.3020 -- 11.3024 \\
		9 & $2\cdot 10^{10}$ & 13.2123 -- 13.2132 \\
		10& $3\cdot 10^{12}$ & 15.1899 -- 15.1901 \\
		\hline
	\end{tabular}
\end{table}

\bigskip

\begin{figure}[h!]
	\includegraphics[width=0.48\textwidth]{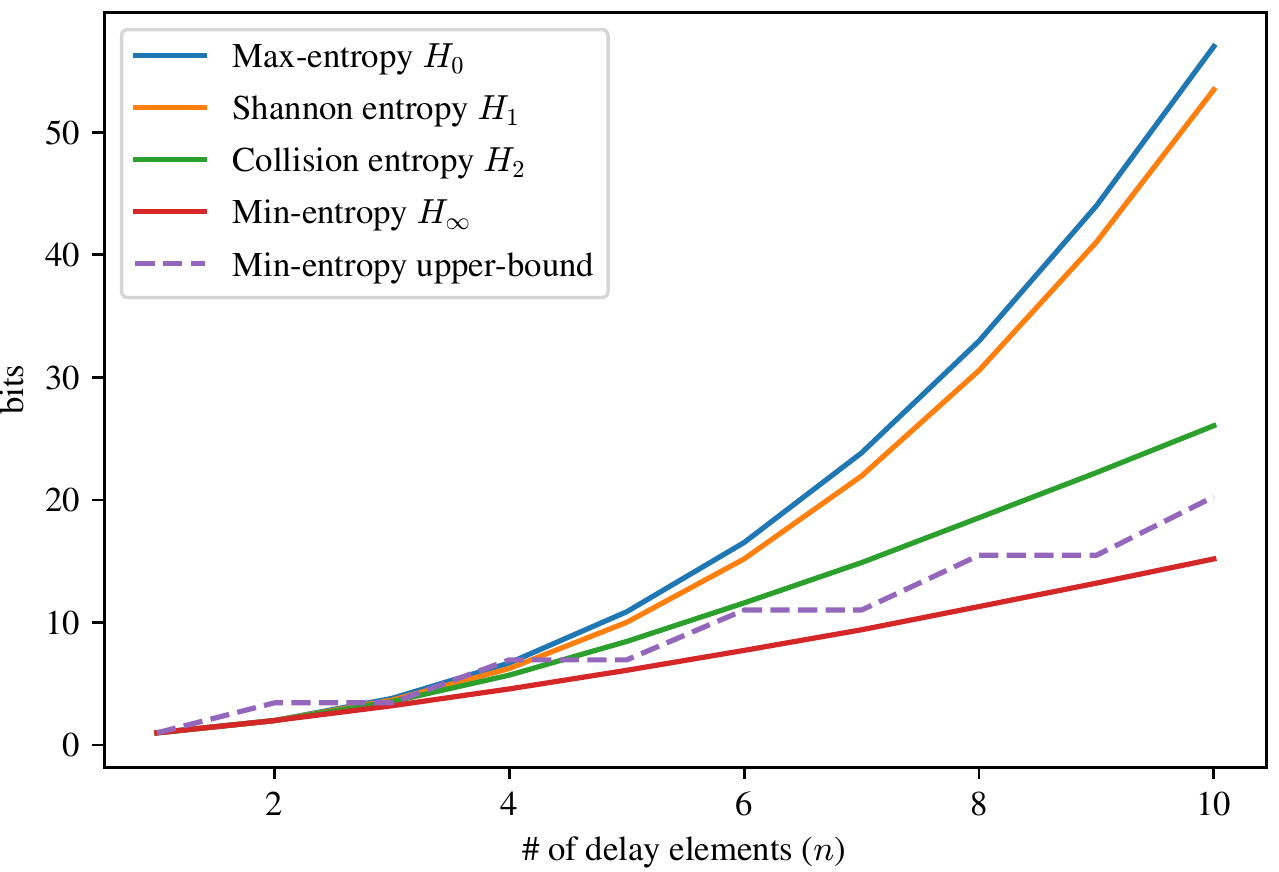}
	\caption{Entropy estimates for $n\leq 10$. The upper bound of the min-entropy (dashed line) is taken from~\cite{delvaux2016upper}.}\label{fig:ent_comp}
\end{figure}

The results of the simulation, up to $n=10$, are presented in Figure~\ref{fig:ent_comp}. The results show that the Shannon entropy is close to the max-entropy, which as seen in Section~\ref{sec:BTF} is  asymptotically equivalent to $n^2$ as $n$ increases.
%
%

\section{Conclusions and perspectives}
\label{sec:conclusion}

While it had been previously shown~\cite{rioul2016entropy} that the entropy of the loop-PUF of $n$ elements could exceed $n$, the exact values were only known for very small values of $n$. Making the link with BTF theory using Chow parameters, we have extended these results to provide accurate approximations up to $n=10$. Our results suggest that the entropy of the loop-PUF might be \emph{quadratic} in $n$: This would be a very positive result for circuit designers, since it implies that the PUF has a very good resistance to machine learning attacks. However, because the min-entropy and collision entropy are much smaller (on the order of $n$) the resistance to cloning may not be as high as expected.

Two interesting theoretical aspects of the PUF entropy are still open: First, to what extent does the entropy of the PUF stay close to the max-entropy for larger values of $n$?  Second, is it possible to obtain a quasi-quadratic entropy in $n$ when choosing a small subset of all $2^n$ possible challenges? The latter point is of great practical interest since it would reduce the time required to obtain the PUF identifier while maintaining a high resistance to machine learning attacks.

For values of $n$ larger than $10$, our method seems to become too costly in space and time to produce accurate estimates of the PUF probability distributions under reasonable conditions. One could perhaps have recourse to entropy estimation methods that dispense with learning the distribution itself, such as the NSB estimation~\cite{nemenman2002entropy}. This could be used to check the predicted trend of the PUF entropy for increasing~$n$.

\section*{Acknowledgment}
The authors would like to thank Prof. Gadiel Seroussi, who first suggested a possible link between our problem and BTF theory at the LAWCI'18 conference in Campinas, Brazil.

\vspace*{0.5cm} 


\bibliographystyle{IEEEtran}
\bibliography{refs}\vspace*{.01pt}

\end{document}